\documentclass[%
 reprint,
 amsmath,amssymb,
 aps,
]{revtex4-1}

\usepackage{graphicx}
\usepackage{dcolumn}
\usepackage{bm}
\usepackage{subfigure}
\usepackage{xcolor}
\usepackage{psfrag}

\begin{document}

\preprint{APS/123-QED}

\def\mean#1{\left< #1 \right>} 
\title{Racetrack Fixed Field Accelerator with pulsed quadrupoles for variable energy extraction}

\author{M. Haj Tahar}
\affiliation{CERN, Geneva, Switzerland} \thanks{Prior affiliation: Columbia University Medical Center}
\author{D.J. Brenner and G. Randers-Pehrson}
\affiliation{%
Center for Radiological Research, Columbia University Medical Center, New York, NY, USA
}%

\begin{abstract}
In this paper, a novel concept combining a fixed field alternating gradient accelerator with time-varying quadrupoles is presented which solves some of the major problems of conventional fixed field accelerators. This concept combines a stable racetrack fixed field configuration with dispersion free straight sections to enable continuous variable energy extraction. In the long straight sections, a time-varying synchrotron-like alternating gradient focusing structure makes it possible for the first time to correct the beam optics of an FFA accelerator in order to heuristically maximize the beam transmission by minimizing the tune excursion and avoiding the crossing of harmful betatron resonances. This accelerator makes it possible to design a compact variable energy extraction machine which is particularly useful for research and medical applications such as heavy ion therapy.  
\end{abstract}

\pacs{Valid PACS appear here}
\maketitle


\section{Introduction}
A Fixed Field alternating gradient Accelerator (FFA) is a concept that was invented in the 1950s independently in the US, Japan and the USSR \cite{symon}\citep{ohkawa}\cite{kolomensky}. Several electron machines were built in the US at the end of the 50s to demonstrate this concept \cite{history}. However, it was not until the 1990s that the interest for FFAs was revived in Japan. One main feature of FFA is that they use fixed field combined function magnets in order to achieve a stable particle motion in the accelerator. Therefore, for heavy ion therapy accelerators requiring energies up to several 100 MeV/nucleon, they are the most viable option for a compact design that fits best into a hospital environment. In addition, with the recent advances in the superconducting magnet technology, the latter has become canonical for many applications. 

To avoid any confusion, FFA can refer to any type of fixed field accelerator. The most worldwide types are cyclotrons  where the main target is to achieve isochronism for all energies thus enabling CW operation. Given that the magnetic field of an FFA can obey any radial or azimuthal dependence, many designs emerged tailored to areas ranging from industrial processes to medical \cite{misu}\cite{garland}\cite{sam} and fundamental research \cite{machida1}\cite{berg}. In the early days, almost all designs were based on highly symmetric rings to reduce the operating restrictions that may arise with systematic field errors. However, recently, several racetrack concepts were proposed \cite{lagrange}. This is generally considered as a solution to practical issues such as the construction cost, the injection and extraction problem, as well as to accommodate and simplify the design of the RF cavity.

Although recently built FFAs demonstrated the concept \cite{aiba}\cite{adachi}, several problems remain.
For instance small gradient field errors lead to a non-scaling of the orbits. The latter contributes to the change of the average horizontal restoring force as well as to the magnetic flutter and induces a variation of the betatron wave number leading to the crossing of transverse resonances during the acceleration cycle. For instance, a gradient field error of the order of $5\%$ yields a variation of the betatron wave number of the order of $7\%$ leading to major beam losses \cite{malek2017}. In addition, one main challenge with optics correction in fixed field accelerators is that the beam moves outward radially during the acceleration. Therefore, a correction scheme should be implemented along the radius of the magnet to produce the desired field profile and is practically difficult since it modifies simultaneously the bending as well as the focusing of the beam. Besides, other errors such as misalignment errors may contribute as well to the beam losses. For medical accelerators, the requirement to achieve an absolute dose accuracy to a few per cent level imposes strong requirements on the stability and reproducibility of the beam parameters \cite{seidel}.

This paper shows that a racetrack FFA configuration can achieve a fixed tune even in presence of field and alignment errors and thus overcome resonance crossings problem. The method is based on creating dispersion free straight sections where a time varying alternating gradient focusing structure is placed. \\

\section{Design principle}
Cyclotrons and linacs are the natural choice for high intensity applications due to their cw operation. However, for cyclotrons, the turn separation for outer orbits becomes small when the beam becomes more relativistic, thus posing a serious beam extraction issue. Nevertheless, several methods exist to overcome this problem: for instance, for the PSI main ring cyclotron, a factor of 3 turn separation is gained at extraction by inducing betatron oscillations around the closed orbit \cite{adelman}. For the TRIUMF cyclotron, $H^{-}$ stripping is employed to extract the beam: by placing several foils at suitable locations in the ring, $H^{-}$ ions are exchanged into protons and extracted at different energies \cite{richardson}. In general, it is considered that 1 GeV is the energy limit for \mbox{cyclotrons \cite{seidel_paper}} which justifies the development of synchrotron and synchro-cyclotron accelerators for heavy ion therapy requiring energies up to 450 MeV/nucleon  \cite{PIMMS} \cite{benedikt}. In this study, one prefers a different class of accelerator, a non isochronous fixed field ring design. In particular fixed field operation allows high repetition rate since the magnets do not need to be ramped up like in a synchrotron. However, one main objective of medical heavy ion accelerators is to target tumors of various depths therefore requiring to tune the beam energy. This is particularly challenging with fixed field accelerators since the beam orbit moves radially outwards during the acceleration so that a continuous beam extraction at the same location is not possible. For this reason, it is common to use a beam degrader instead. This reduces the energy of the output beam but also reduces its quality due to multiple scattering. Thus, activation due to beam losses is an issue that requires more shielding of the accelerator. Besides, the beam intensity becomes strongly dependent on the beam energy which complicates the heavy ion treatment \cite{schippers} and justifies once again the wide use of synchrotrons for this application.

In order to remediate this problem and continuously extract the beam at various energies, the idea is to start with the design of a fixed tune scaling FFA and modify the concept in order to create dispersion free straight sections.

The design approach is based on solving the Lorentz equation of motion in simulated magnetic fields and adjusting the magnet field profiles to achieve stable trajectories in both horizontal and vertical planes. The beam optics calculation are based on the particle tracking code ZGOUBI \cite{zgoubi} which solves the non-linear equation of motion using truncated Taylor expansions of the field and its derivatives up to the $5th$ order. A median plane field map is first generated and extrapolation out of it is achieved by means of Taylor series. For this, the median plane symmetry is assumed and the Maxwell equations are accommodated. This yields results in good agreement with the 3D field map simulations. \\ 

\subsection{Scaling FFA}
\begin{figure}[htb]
\centering 
\includegraphics*[width=9cm]{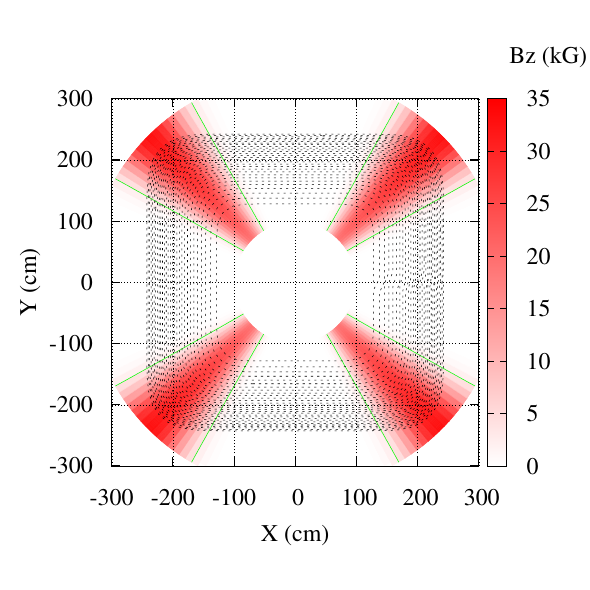}
\caption{Vertical component of the magnetic field in the median plane of the scaling FFA with some of the closed orbits between injection and extraction energies (shown in dashed black curves). In this configuration, the average field index of the magnet is $k = 0.45$.}
\label{scaling1}
\end{figure}

\begin{figure}[htb]
\centering 
\includegraphics*[width=7cm]{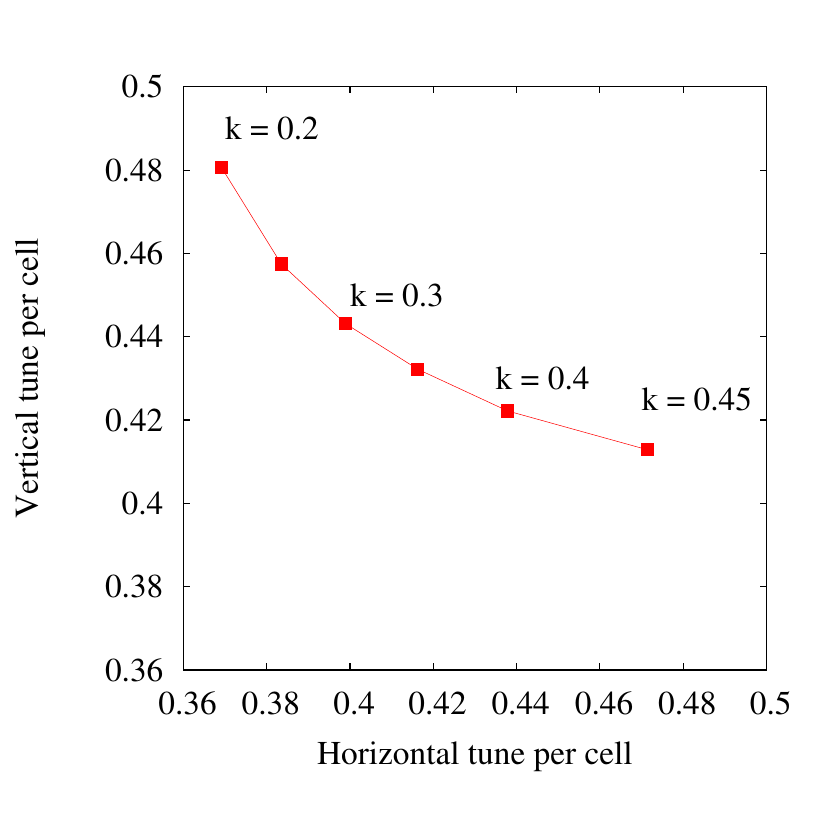}
\caption{Evolution of the horizontal and vertical tunes per cell as a function of the average field index of the magnet: on the right side one is limited by the radial $\pi$-mode stop-band resonance, while on the left side the limiting factor is the vertical $\pi$-mode stop-band resonance.}
\label{scaling}
\end{figure}

A scaling FFA is chosen as a first step in order to achieve a stable lattice with no crossing of the transverse betatron resonances. This ensures a large dynamic acceptance which is particularly important for high power applications and isotope production. As illustrated in fig \ref{scaling1}, a configuration with 4 identical radial sector magnets is preferred since it provides enough space between the magnets to place injection/extraction devices as well as other elements discussed below. In order to have the most compact design, no reverse bend is considered. 
The field profile for each magnet writes as a separable function in radial and azimuthal coordinates in the following way: $B(R,\theta)=\mathcal{R}(R)F(\theta)$ where $B$ is the vertical component of the magnetic field in the median plane, $R$ is the radial coordinate with respect to the center of the ring and $\theta$ is the azimuthal coordinate. The fringe field falloff of the magnets is represented by the Enge model \cite{enge} to describe the azimuthal field variations, i.e. the flutter function $F$. The latter writes in the following way
\begin{eqnarray}
F(\theta) = \dfrac{1}{1+\exp\left[P_1(\theta)\right]} \times \dfrac{1}{1+\exp\left[P_2(\theta)\right]}
\end{eqnarray}
where the subscripts 1,2 denote the entrance, exit of the magnet and the polynomials $P_i$ are defined by:
\begin{eqnarray}
P_i(\theta) = C_{i0}  &+& C_{i1} \left(\theta - \theta_i \right) + C_{i2} \left(\theta - \theta_i \right)^2  \nonumber \\
                      &+& C_{i3} \left(\theta - \theta_i \right)^3
\end{eqnarray} 
The radial field profile obeys the law of a scaling FFA, i.e. $\mathcal{R}(R)=B_0 (R/R_0)^k$ where $k$ is the average field index of the magnet and $B_0$ is the field at $R=R_0$. This guarantees a large transverse dynamic acceptance. Given that the horizontal tune per cell is limited by the radial $\pi$-mode stop-band resonance, this sets a limit on the average field index of the magnet that can be chosen to achieve stable transverse motion. \mbox{Fig \ref{scaling}} shows the evolution of the vertical and horizontal tunes as a function of the $k$-value. Qualitatively, one can see that increasing the $k$-value increases the horizontal tune and decreases the vertical one as predicted by the Symon formula for a radial sector machine \cite{symon}:
\begin{eqnarray}
\nu_x ^2 &\approx & k+1 \\
\nu_y ^2 &\approx & -k + \mathcal{F}^2
\end{eqnarray}
where $\mathcal{F}$ is the magnetic flutter defined by:
\begin{eqnarray}
\mathcal{F}^2 = \dfrac{\mean{B^2}-\mean{B}^2}{\mean{B}^2} = \dfrac{\mean{F^2}-\mean{F}^2}{\mean{F}^2}
\end{eqnarray}
However, the above equations are based on the smooth approximation which consists on taking the averages of the equation of motion and fail when the number of sectors is small and when there are azimuthal field errors leading to a non-scaling of the orbits \cite{malek2017}. 
 \\
Next, the idea is to transform the 4-sector radial FFA into a racetrack configuration with dispersion free straight sections. Historically, the first idea to insert dispersion-free long straight sections for FFA was introduced by \mbox{P. F. Meads \cite{meads1} \cite{meads2}.} Later on, it was shown that with a transverse exponential magnetic field law, straight sections in scaling FFA can be designed where the dispersion functions are considerably reduced \cite{lagrange0}. 
To reduce the radial orbit excursion at locations where the RF cavities can be installed, T. Planche et al introduced the idea to combine scaling FFA cells with different field indices \cite{planche}.

\subsection{Racetrack FFA}
\begin{figure}[htb]
\centering 
\includegraphics*[width=9cm]{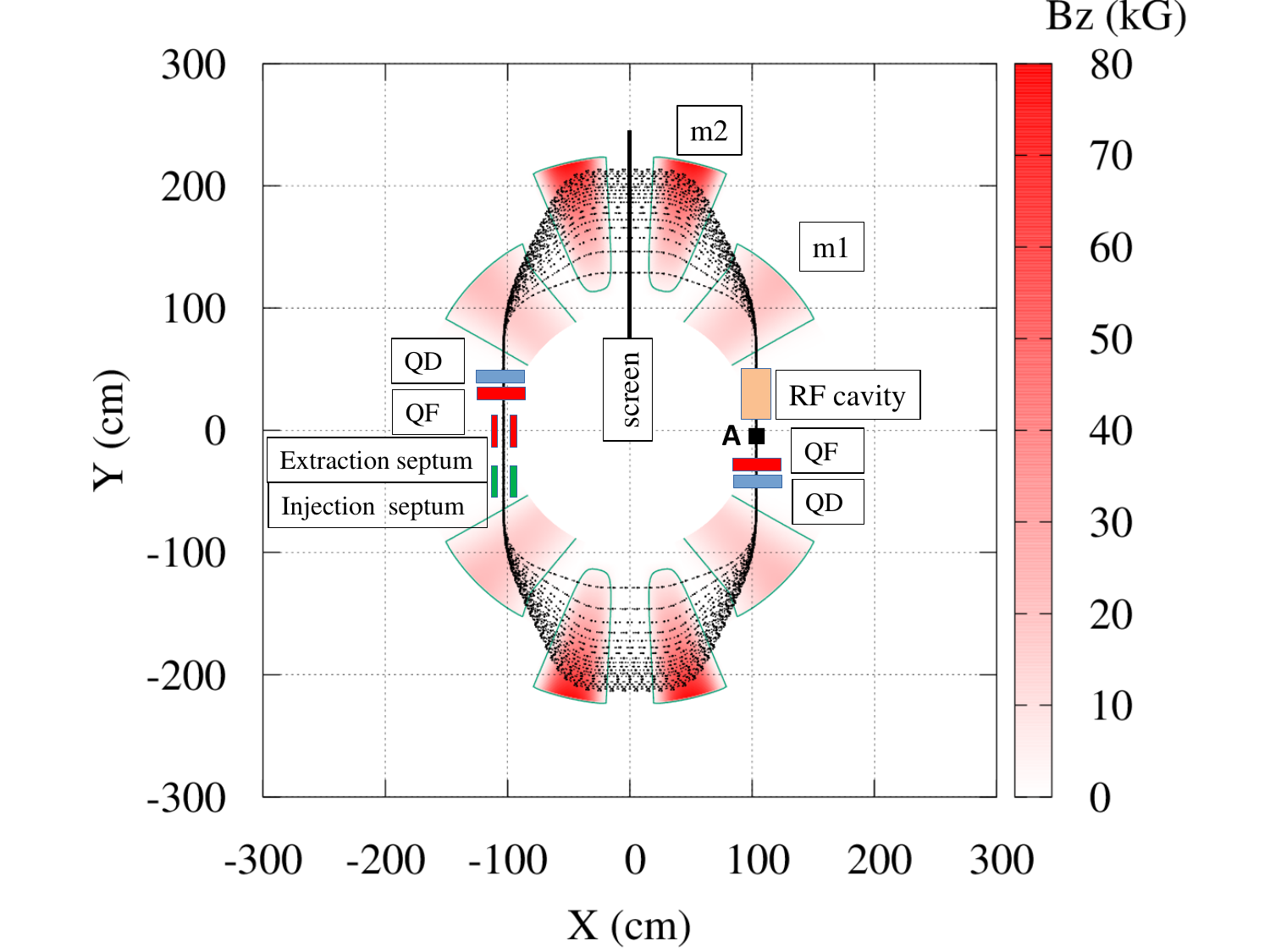}
\caption{Vertical component of the magnetic field in the median plane of the accelerator with some of the closed orbits between injection and extraction energies (shown in dashed black curves). QF and QD refer to the focusing and defocusing trim quadrupoles respectively.}
\label{racetrack}
\end{figure}

\begin{figure}[htb]
\centering 
\includegraphics*[width=9cm]{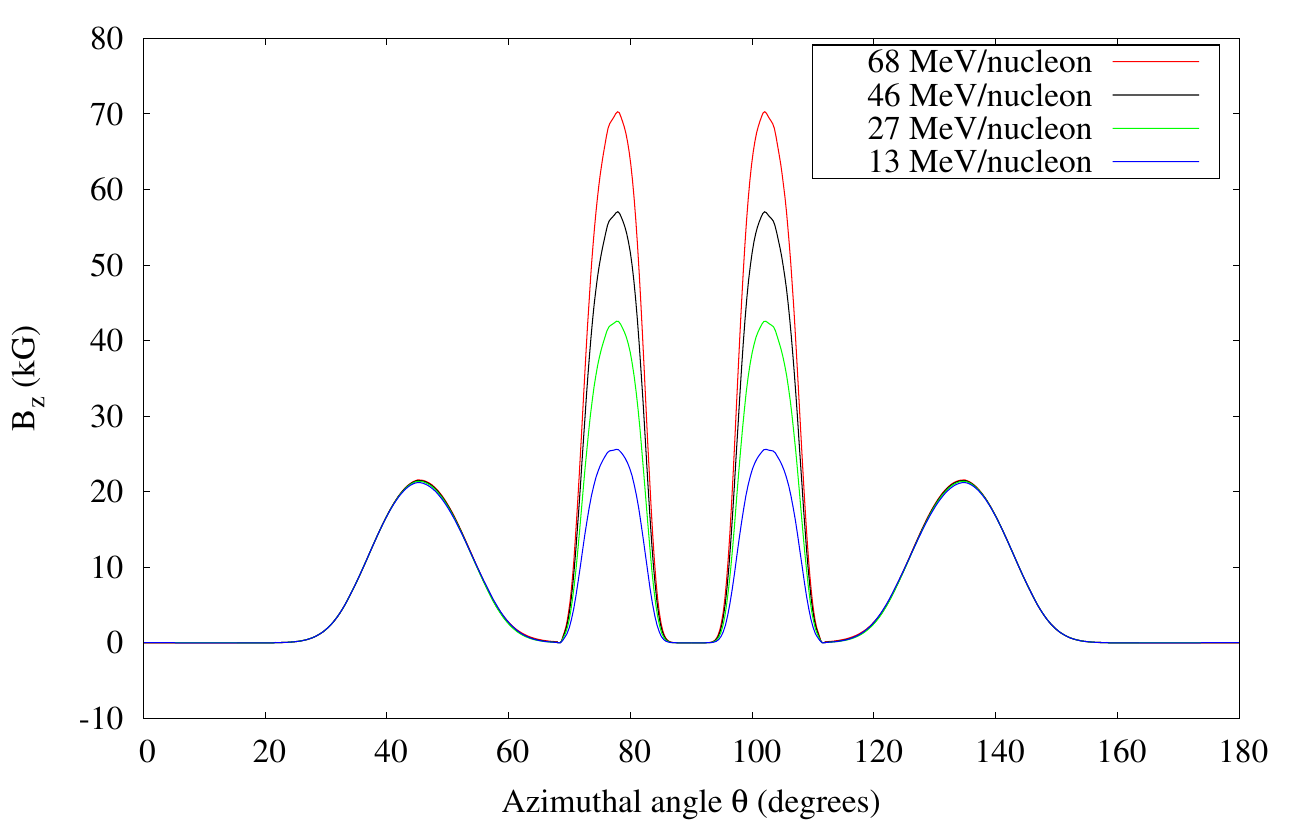}
\caption{Magnetic field along several particle trajectories.}
\label{magnetic_field}
\end{figure}

\begin{figure}[htb]
\centering 
\includegraphics*[width=9cm]{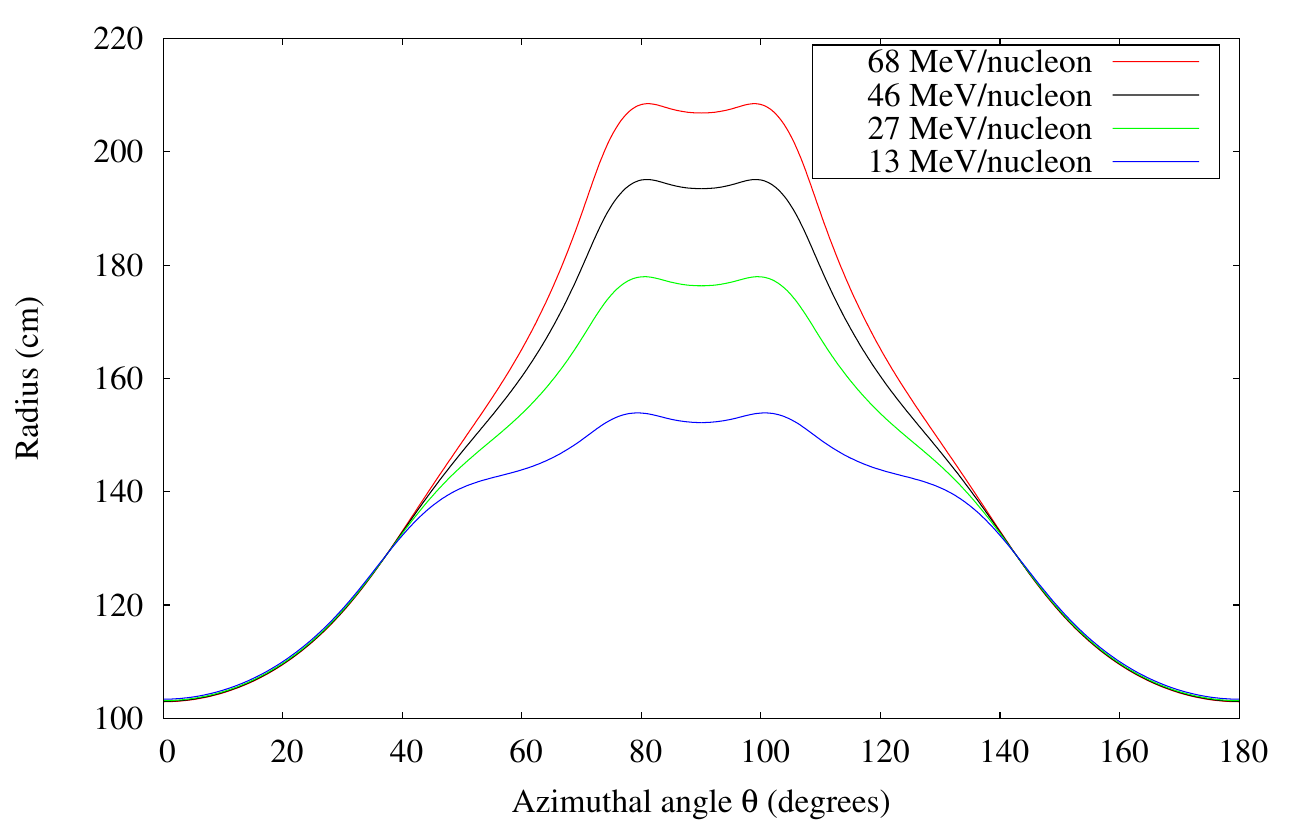}
\caption{Closed orbit trajectories at various energies.}
\label{traj}
\end{figure}

In our approach, a racetrack configuration is achieved by replacing the 4-fold symmetry with a 2-fold symmetry machine: the main idea is to add a second magnet (m2) for each sector and optimize its field profile in order for the pair (m1+m2) to create a 90 degree bending angle for all energies. This is achieved using the beam dynamics tracking code ZGOUBI \cite{zgoubi}: a fitting method is employed which consists in tracking particles with different momenta and the same initial coordinates (point A in fig \ref{racetrack}) to the screen where their final coordinates are recorded. The field along the radius of magnet (m2) is adjusted until all particles exit perpendicular to the screen. Once this is achieved, the ring is completed by creating mirror symmetries with respect to the screen and to the line passing by point A and perpendicular to the screen. A microtron-like racetrack FFA configuration with dispersion-free straight sections is thus achieved as can be seen in fig \ref{racetrack}. The magnetic field along several particle trajectories is shown in fig \ref{magnetic_field}. In magnet (m1), the field experienced for various energies is almost the same given that the average field index is small. In magnet (m2), the orbits are well separated (see fig \ref{traj}) and the field increases rapidly in order to provide the \mbox{90 degrees} bending angle for all energies.

The advantages of the dispersion-free sections are numerous: for instance, the straight section can be used to dump the beam after tripping during the acceleration which helps reduce the activation of the accelerator components. This is a common problem for cyclotrons for instance. In addition, tight space for the RF cavity causes coupling with the leakage field of the main lattice magnets leading to closed orbit distortion issues \cite{cod}. Placing the RF cavity in a long field-free straight section  remediate such issues. \\
Next, one investigates the stability of the particle trajectories in the racetrack FFA concept.

\section{Beam stability analysis}
\subsection{Betatron oscillations}
\begin{figure}[htb]
\centering 
\includegraphics*[width=7cm]{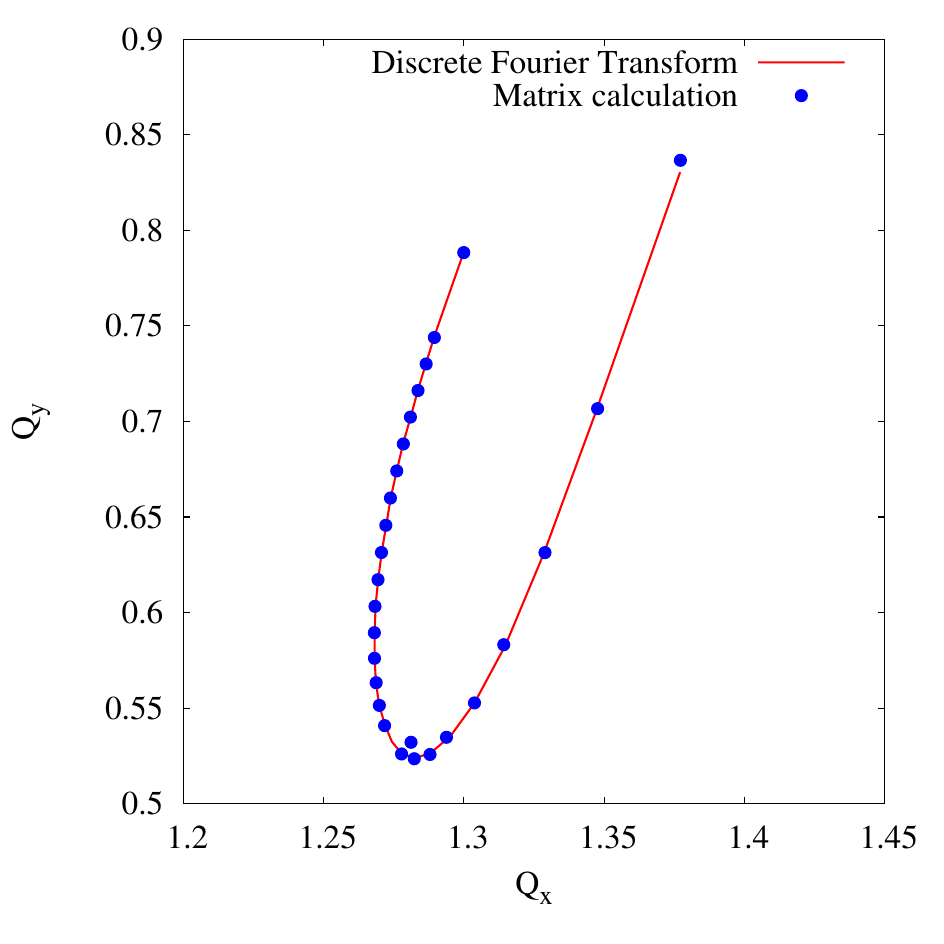}
\caption{Betatron tunes from 12 to 75 MeV/nucleon (right to left).}
\label{tune0}
\end{figure}

\begin{figure}[htb]
\centering 
\includegraphics*[width=9cm]{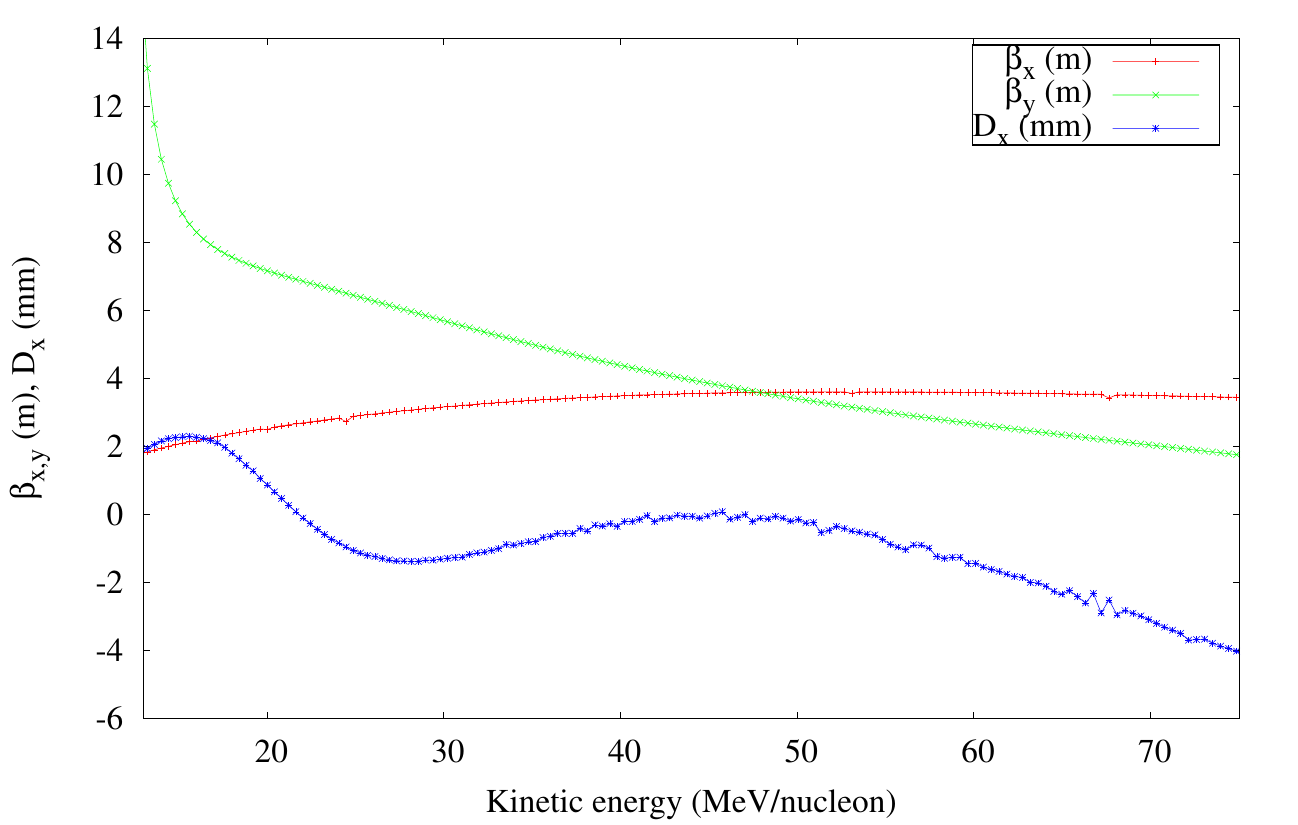}
\caption{Periodic betatron functions from 12 to 75 MeV/nucleon at the center of the straight section (point A). The periodic dispersion functions are shown in blue (units of mm).}
\label{beta}
\end{figure}

The stability analysis is part of the optimization process: the first step is to adjust the edge angle as well as the average field index $k$ of (m1) in order to ensure enough strong focusing and avoid the crossing of integer and half integer resonances. The main idea is that the bare lattice, i.e. without any added time-varying elements, be stable in order to characterize its beam optics. As shown in \cite{johnstone}, a tune-stabilised lattice can be achieved by employing the edge focusing which occurs when the particle orbit is non perpendicular to the entrance and/or exit faces of the magnet. In the current design, only the exit edge angle of (m1) is optimized since it has the largest impact on reducing the tune excursion. The entrance edge angle of (m1) has a limited impact on the tunes since all incoming orbits overlap at that location. The parameters of the test lattice are listed in Table \ref{tab:table1}:
\begin{table}
\caption{\label{tab:table1}Main parameters of the test lattice.}
\begin{ruledtabular}
\begin{tabular}{lcr}
Parameter&Value&Unit\\
\hline
Charge to mass number ratio & 1/2 & -\\
Circumference & 15 & m\\
Length of field-free straight section & 1 & m\\
Field index $k$ of m1 & 0.75 & - \\
Injection/extraction kinetic energy & 12/75 & MeV/nucleon\\
Maximum field & 8 & T \\
\end{tabular}
\end{ruledtabular}
\end{table}
In summary, the transverse tunes of the optimized lattice are shown in \mbox{fig \ref{tune0}} where one can see that the integer and half integer resonances are avoided. The effect of the edge angle is mainly dominant for the higher energies (left hand side of the plot). The tunes were computed by performing a discrete Fourier transform on the particle coordinates stored turn after turn and choosing the maximum amplitude from that to be the cell tune. \mbox{A second} approach consists in computing the transfer matrices at various energies by tracking paraxial rays around each closed orbit. The tunes are then determined from the trace of the matrix and a comparison of both results is shown in \mbox{fig \ref{tune0}.}
Next, the periodic lattice functions as a function of the energy are computed at the location of the straight sections (point A in fig \ref{racetrack}). As shown in \mbox{fig \ref{beta}}, the lattice functions vary smoothly with the energy and the dispersion function is reduced to the mm level. 
 
\subsection{Phase space trajectories at fixed energies} 
\begin{figure}%
\centering
\subfigure[][ $E_{kin} = 13 \mbox{MeV/nucleon}$. $\epsilon_x \approx 100$ mm.mrad]{%
\label{fig:ex3-a}%
\includegraphics[height=2in]{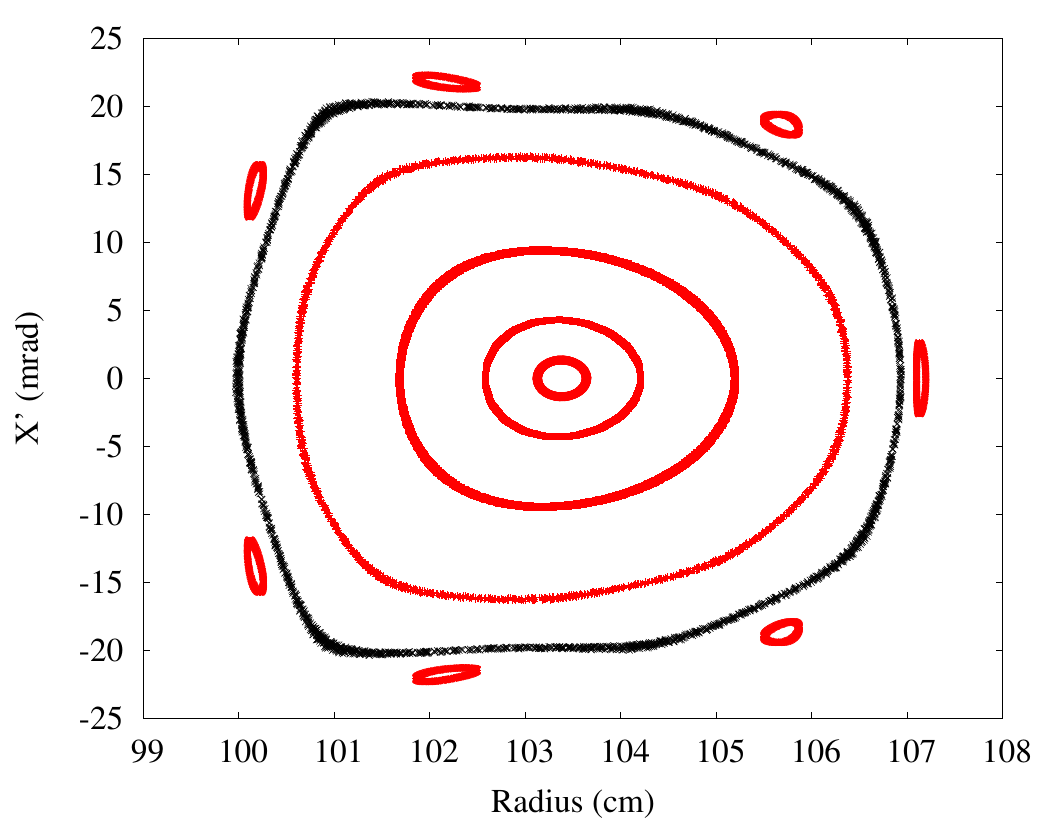}}%
\hspace{8pt}%
\subfigure[][ $E_{kin} = 27 \mbox{MeV/nucleon}$. $\epsilon_x \approx 165$ mm.mrad]{%
\label{fig:ex3-b}%
\includegraphics[height=2in]{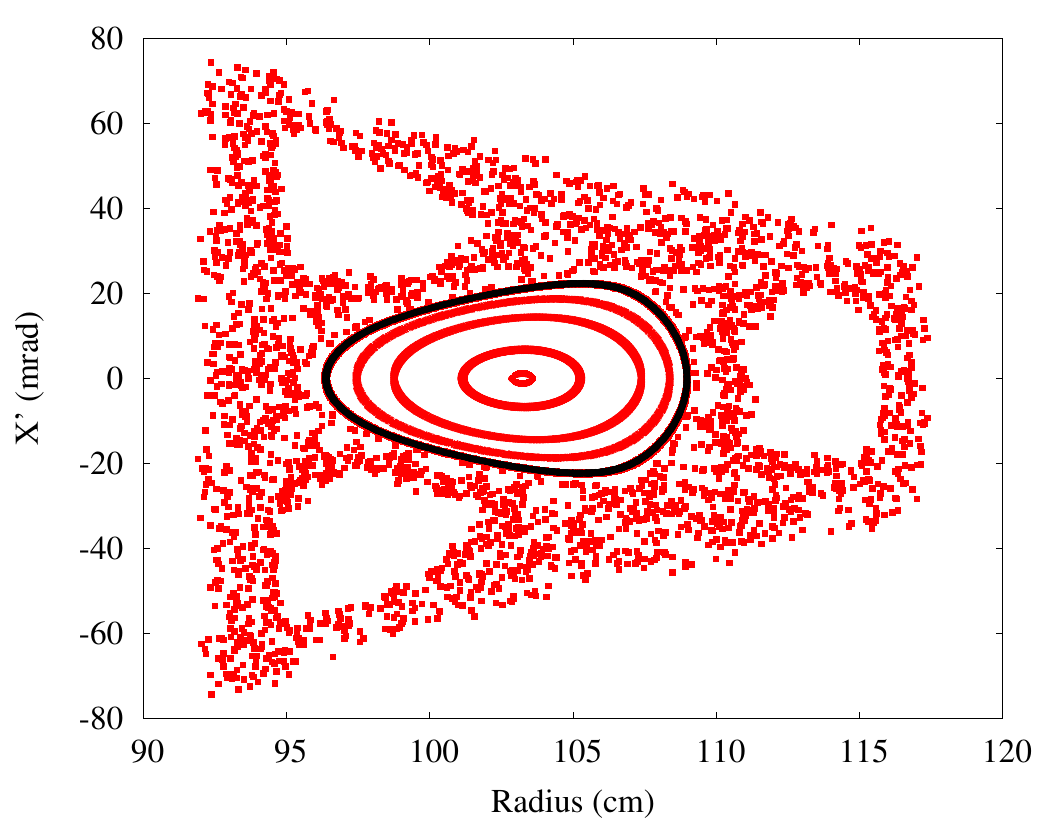}} \\
\subfigure[][ $E_{kin} = 46 \mbox{MeV/nucleon}$. $\epsilon_x \approx 430$ mm.mrad]{%
\label{fig:ex3-c}%
\includegraphics[height=2in]{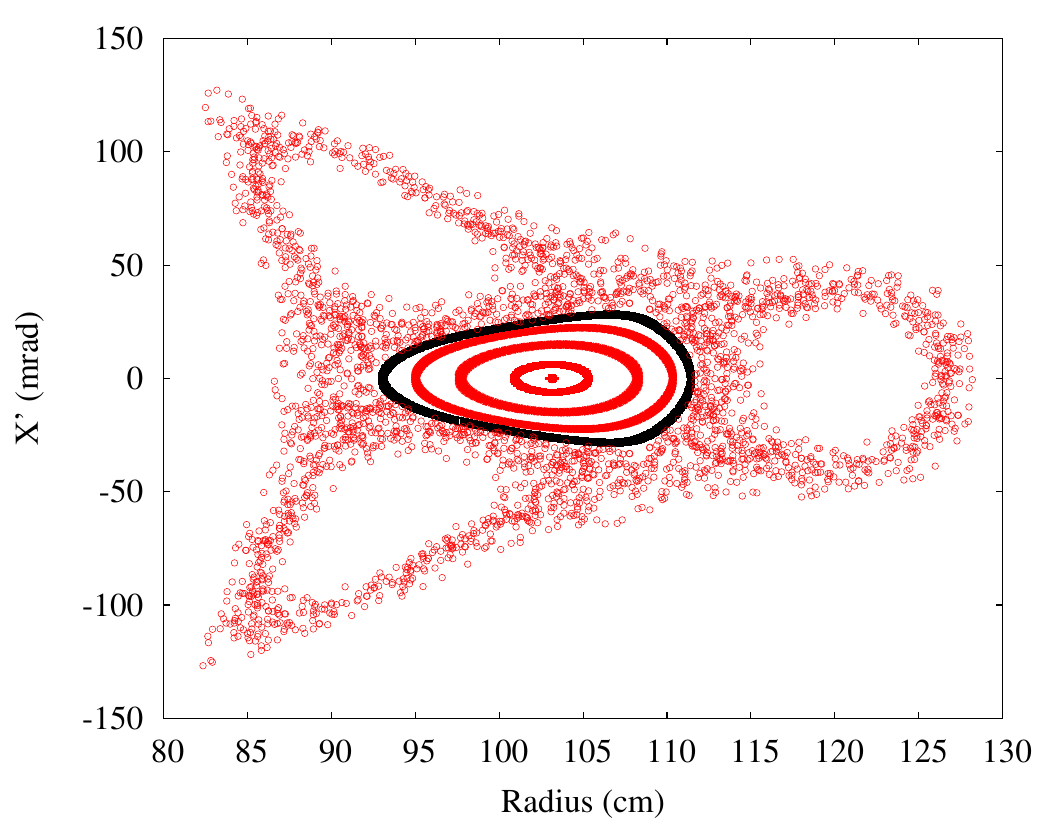}}%
\hspace{8pt}%
\subfigure[][ $E_{kin} = 68 \mbox{MeV/nucleon}$. $\epsilon_x \approx 340$ mm.mrad]{%
\label{fig:ex3-d}%
\includegraphics[height=2in]{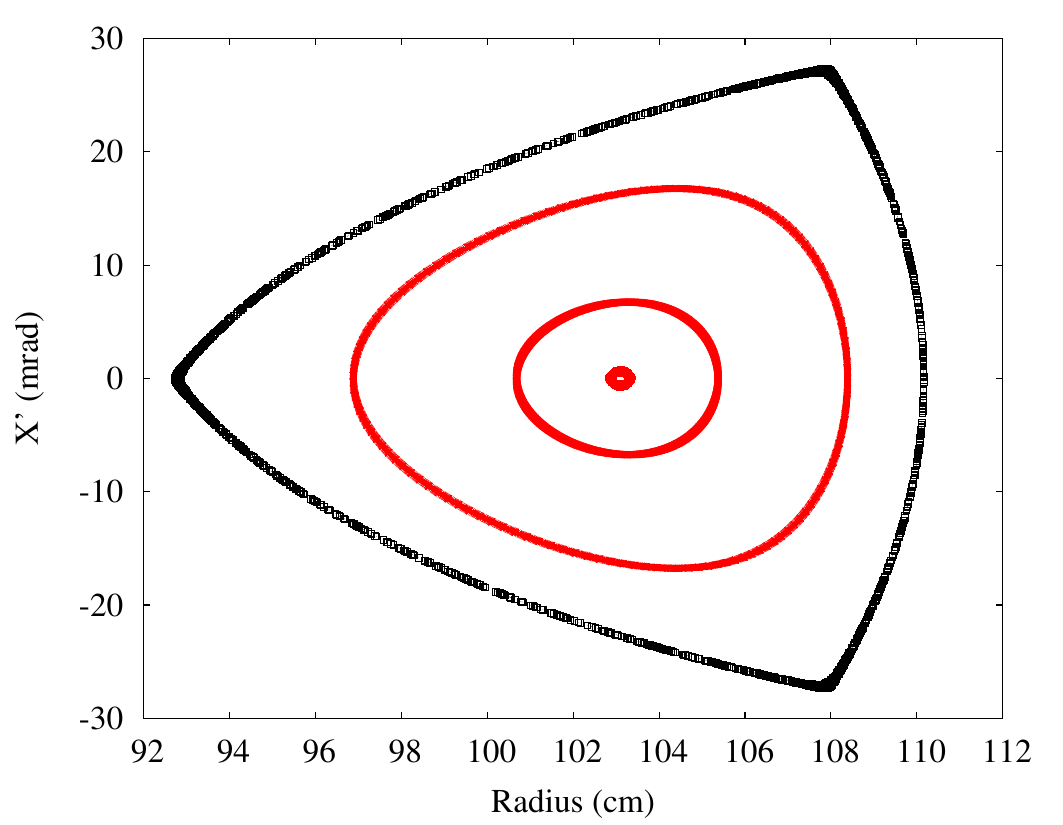}}%
\caption[A set of four subfigure.]
{Phase space trajectories at different energies, including the separatrix (shown in black). The vertical beam emittance considered for all cases was limited to less than \mbox{1 mm.mrad.}
}%
\label{fig:da_fixed}%
\end{figure}

The optics of the racetrack FFA are highly non-linear. Therefore, there is a tune dependence with the amplitude of the displacement from the closed orbits. This can provoke resonant phenomena in the phase space leading to the growth of the particle oscillations and to major beam losses. For this reason, it is crucial to assess the long term behavior of the particle trajectories. As a first approach, one evaluated the Dynamic Aperture (DA) at fixed energies by performing 2000 turns of particle tracking around different closed orbits. The DA is defined as the maximum normalized emittance that the beam can have without losses due to single particle dynamics effects. We define it from the largest stable horizontal displacement around the closed orbit at point A in the following way:
\begin{eqnarray}
\epsilon_x = \dfrac{\beta \gamma}{\beta_x} x^2
\end{eqnarray}
Fig \ref{fig:da_fixed} shows the horizontal phase space trajectories at different energies, including the separatrix. Large beam sizes can be accommodated. Due to third integer resonances, the phase space trajectories for large amplitude particles is distorted from elliptical to triangular.  
 
\subsection{Trim quadrupoles}
The trim quadrupoles aim to reduce the tune excursion in order to avoid the crossing of any type of resonance, up to the 3rd order, known to be detrimental to the beam. To achieve this, one introduces time-varying quadrupole magnets in the long straight sections.
To begin with, one constructs the one turn transfer map of the bare lattice. This is achieved by tracking particles with small displacements from each closed orbit, i.e. each different particle energy. One way to validate the result is to compare the tune calculation using two different approaches as shown in fig \ref{tune0}. Then, one adds the trim quadrupoles QF and QD in a way that maintains the 2-fold symmetry of the accelerator as shown in \mbox{fig \ref{racetrack}.} This is of particular importance in order not to increase the number of betatron resonances that may be driven. In our case, one considers 4 quadrupole doublets (QF,QD) each quadrupole having a length of $0.1 \text{ m}$.  The transfer map of the time-varying lattice is thus re-calculated using a MATHEMATICA code and the gradients of the trim quadrupoles are determined for each particle energy in order to match the tunes of a chosen working point. For instance, one can compute the stability diagram of the lattice as a function of the gradients of the quadrupole magnets as shown in \mbox{fig \ref{stability}} below: increasing the gradient of the focusing (resp. defocusing) quadrupoles increases (resp. decreases) the horizontal tune and decreases (resp. increases) the vertical one. Furthermore, increasing the gradients of the focusing and defocusing quadrupoles simultaneously results in increasing the tunes in both planes. This is the principle of strong focusing accelerators \cite{courant}. 
Due to the existence of resonances, the stability diagram is crossed by several bands in the neighborhood of the lines $Q_{x,y} = k/N$, $k$ and $N$ being integers and $N$ is the order of the resonance (coupling resonances are not represented here). If one requires that the betatron frequencies should not take their resonance values, then the stability diagram is divided into individual cells inside which the operating point must be chosen. As a first step, one restricted the total tune variations to within half an integer. As a next step, one can require that the tunes remain above both third integer resonances $3Q_x=4$ and $3Q_y=2$. This sets the limits on the gradients of the trim quadrupoles (lightblue region). \\
To the first order, one can represent the effect of the trim quadrupoles as a localized gradient field error in the lattice. Therefore, the resulting ring tune shift for each trim quadrupole can be approximated \mbox{by \cite{courant}:}
\begin{eqnarray}
\Delta \nu_x = \dfrac{1}{4\pi} \int_0^C \beta_x(s) K_x(s) ds \approx \dfrac{\mean{\beta_x}_q G L_q}{4 \pi B\rho}
\end{eqnarray}
where $\mean{\beta_x}_q$ is the average betatron function for a quadrupole of length $L_q$ and field strength $G$. For a kinetic energy of 43 MeV/nucleon where $\mean{\beta_x}_q \approx 3.5 \text{ m}$ and $\mean{\beta_y}_q \approx 4 \text{ m}$, quadrupole gradients of \mbox{1.7 T/m} and \mbox{1.5 T/m} are sufficient to shift the tunes by 0.1 in the horizontal and vertical planes respectively. This is in agreement with the matrix calculations. 
\begin{figure}[htb]
\centering 
\includegraphics*[width=7cm]{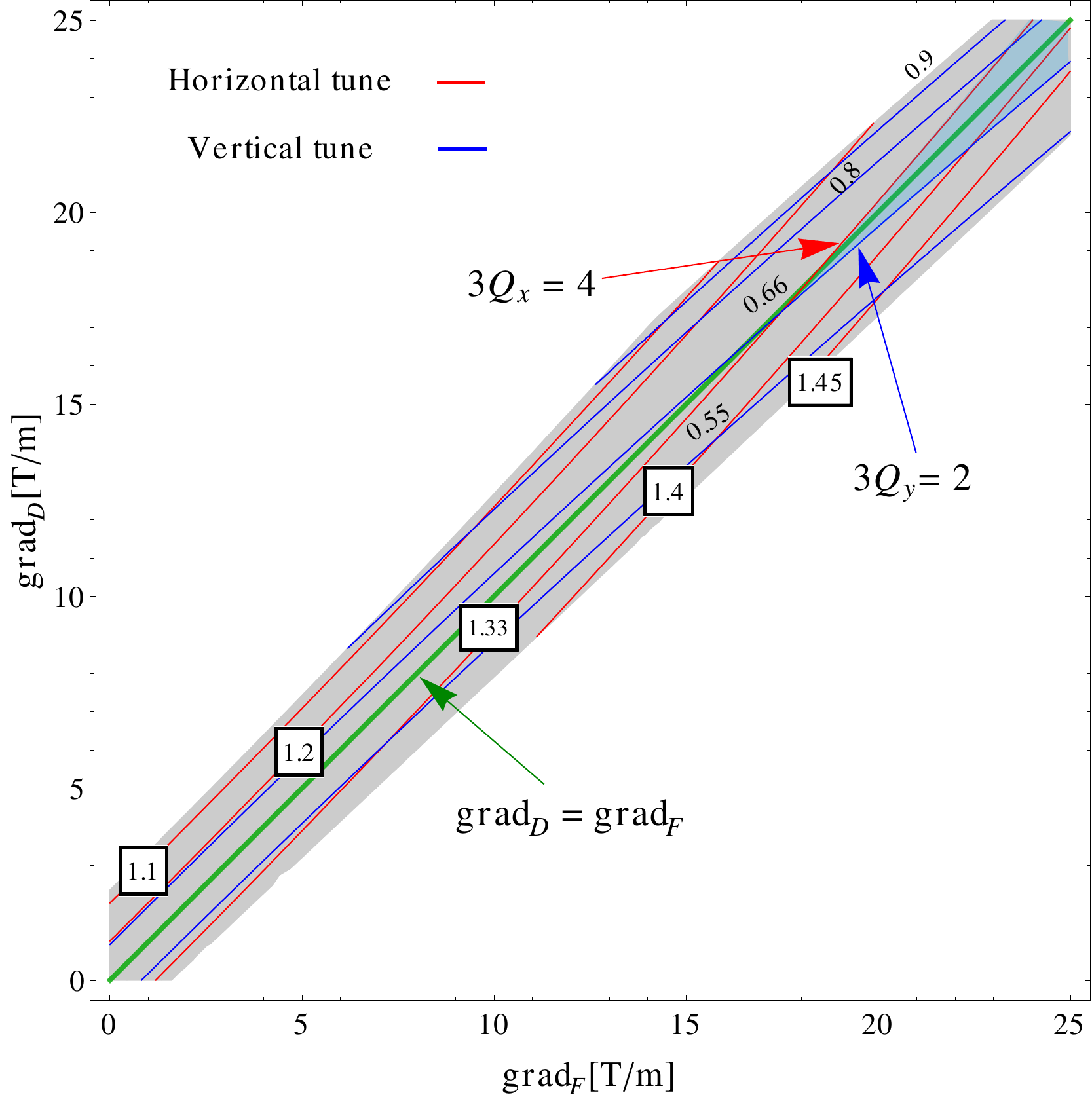}
\caption{Stability diagram of the racetrack FFA as a function of the gradients of the Focusing (F) and Defocusing (D) time varying quadrupoles at a given kinetic energy of 43MeV/nucleon. The tunes of the bare lattice are \mbox{$(Q_x,Q_y)=(1.26,0.60)$.}}
\label{stability}
\end{figure}

\begin{figure}[htb]
\centering 
\includegraphics*[width=7cm]{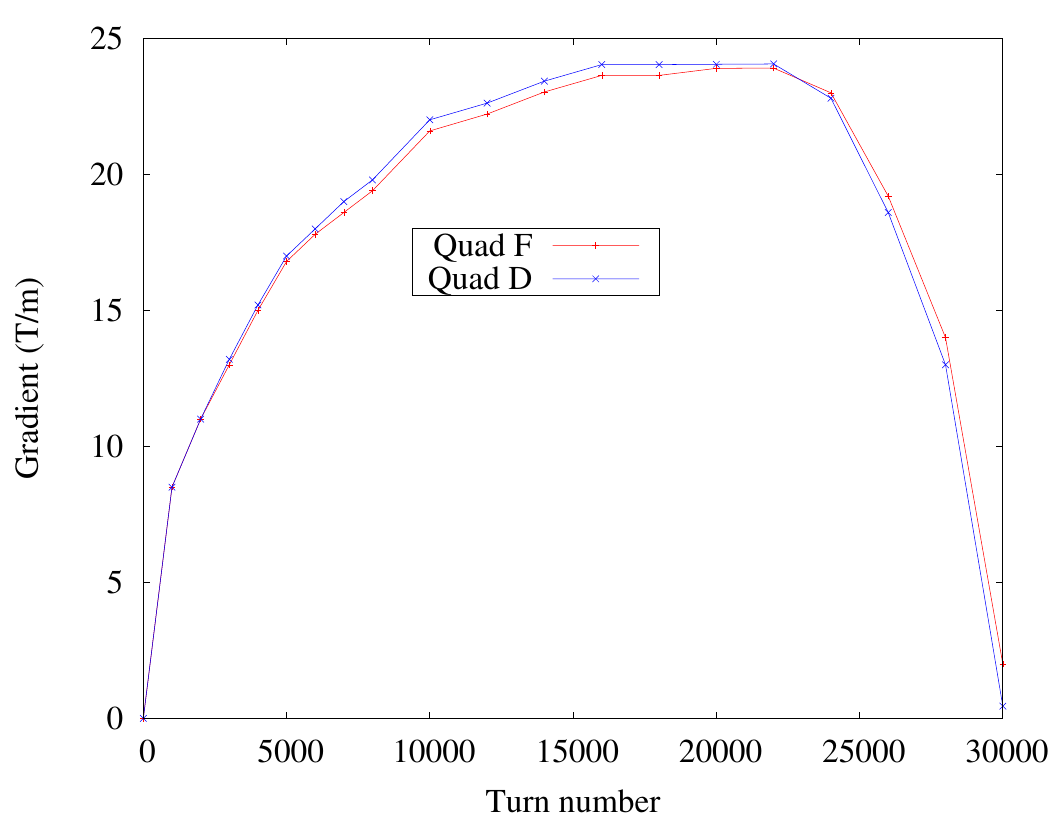}
\caption{Gradient of the magnetic field of the focusing and defocusing trim quadrupoles as a function of the turn number.}
\label{ramping}
\end{figure}

\begin{figure}[htb]
\centering 
\includegraphics*[width=7cm]{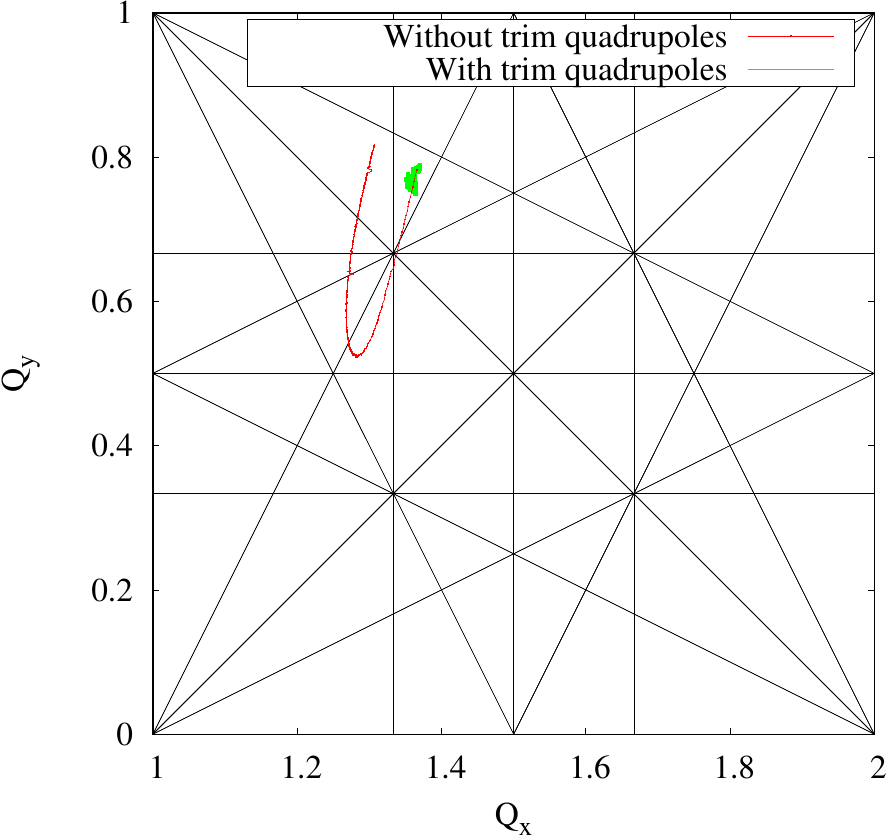}
\caption{Betatron tunes from 12 to 75 MeV/nucleon (right to left). The solid lines in black show the resonance lines up to the 3rd order.}
\label{tune}
\end{figure}

The solution obtained is finally implemented in the tracking code ZGOUBI by adding the time-varying elements to the bare lattice where one re-scaled the gradient strength on a turn by turn basis as shown in fig \ref{ramping}. The results of the tune calculation from the accelerated orbit is shown in green in fig \ref{tune} where one proved that the resonance crossing problem is overcome. In summary, the time varying field of the trim quadrupoles compensates for the monotonic behavior of the tunes in the bare lattice. For instance, increasing simultaneously the gradients of QF and QD allows to increase both horizontal and vertical tunes. To our knowledge, this is the first time that such a property is established in a non-scaling FFA \cite{patent_Malek}. In addition, the merit of the described method is that it applies to any bare lattice in the presence of field or misalignment errors of its main magnets: practically, the working point can be optimized in a way to heuristically maximize the overall beam transmission. This is of utmost importance for the study of the resonance crossing problem in fixed field accelerators.

\subsection{Accelerated orbit analysis}
\begin{figure}[htb]
\centering 
\includegraphics*[width=7cm]{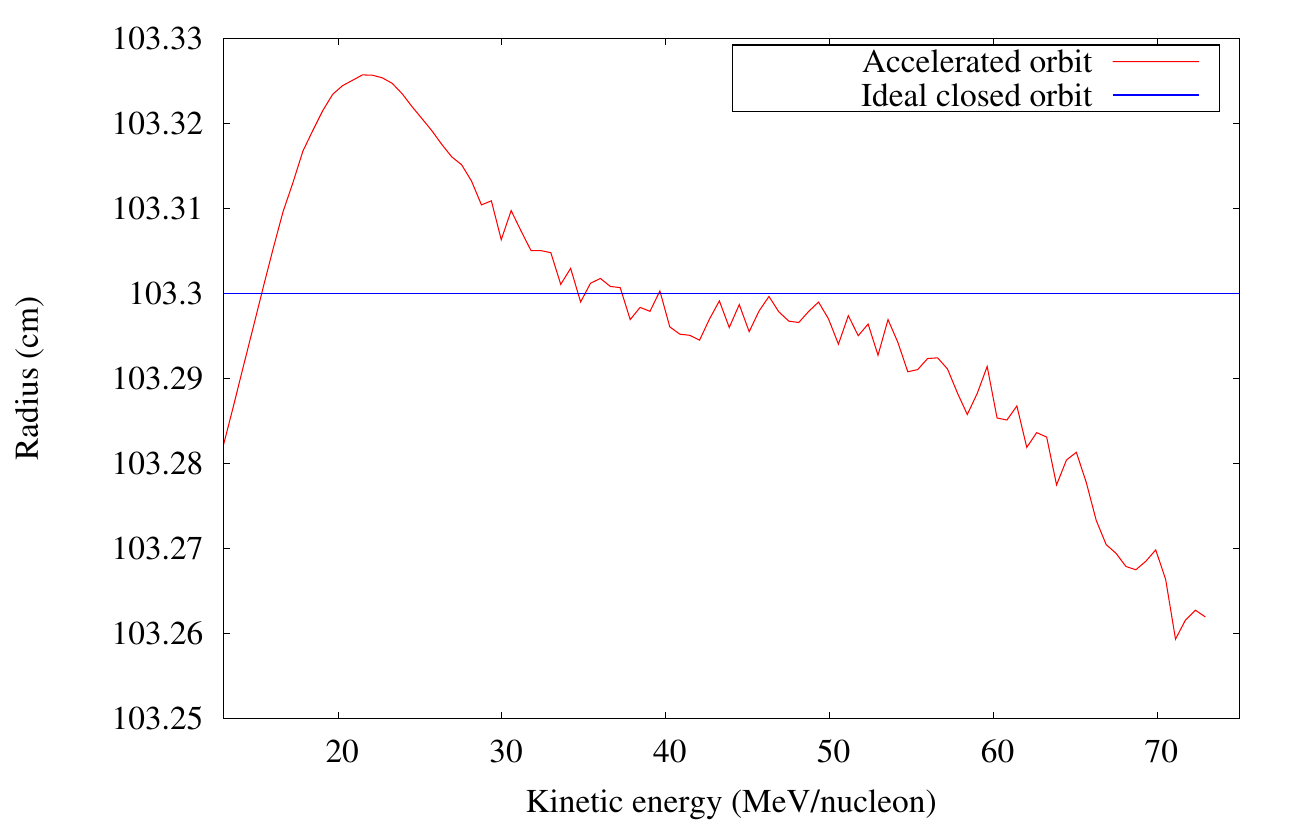}
\caption{Accelerated orbit as a function of the energy \mbox{(30000 turns)} at the location of the straight sections: the oscillation around the closed orbit is around $\pm 300 \text{ $\mu$m}$.}
\label{accel_or}
\end{figure}

\begin{figure}[htb]
\centering 
\includegraphics*[width=8cm]{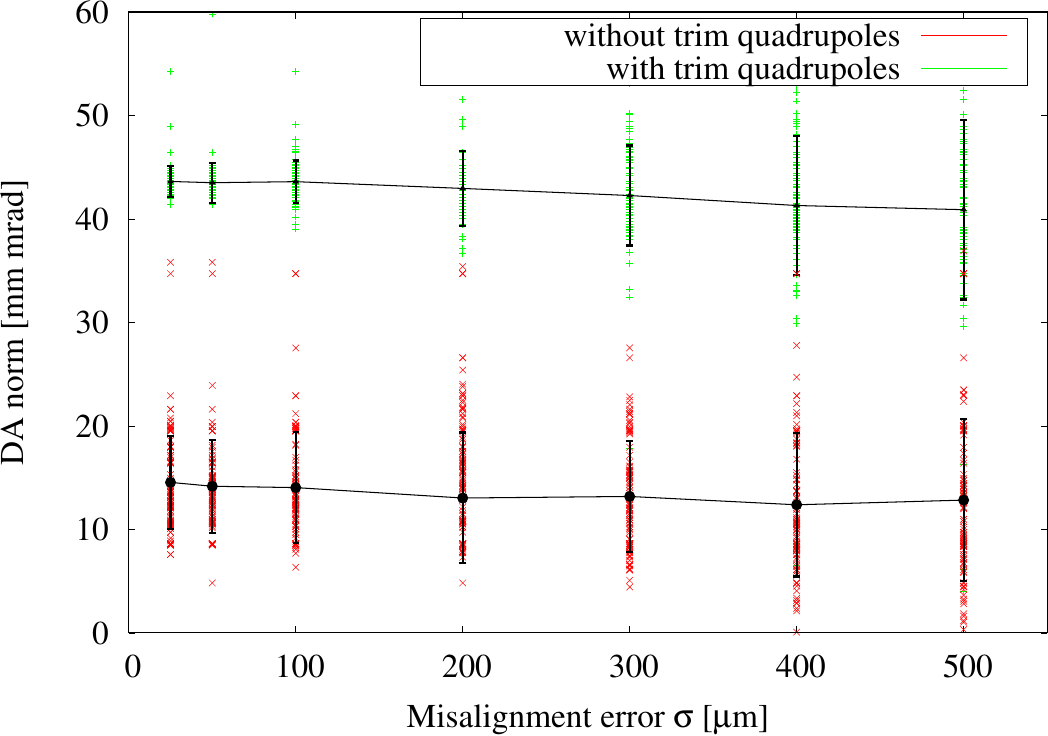}
\caption{The DA as a function of the horizontal and vertical cell misalignment errors with a Gaussian distribution of \mbox{width $\sigma$.}}
\label{da}
\end{figure}

The previous formalism which consists in calculating local quantities such as the betatron functions or the phase space trajectories in the vicinity of different closed orbits corresponding to different energies is only valid if all quantities evolve adiabatically with time. For instance, in presence of resonance crossing problem and/or non matched optics, the beam emittance can increase and particle losses may occur, thus justifying the implementation of the trim quadrupoles. Therefore, the next step in the simulations is to track particles throughout the acceleration cycle by introducing magnet misalignments in order to characterize their stability in presence of imperfections. In the racetrack FFA, the orbit of the synchronous particle inside the main magnets spirals outwards with increasing energy. Therefore, the RF phase of the accelerating cavity must evolve in such a way as to follow the change of the revolution frequency of the particle. 
\begin{figure}[htb]
\centering 
\includegraphics*[width=9cm]{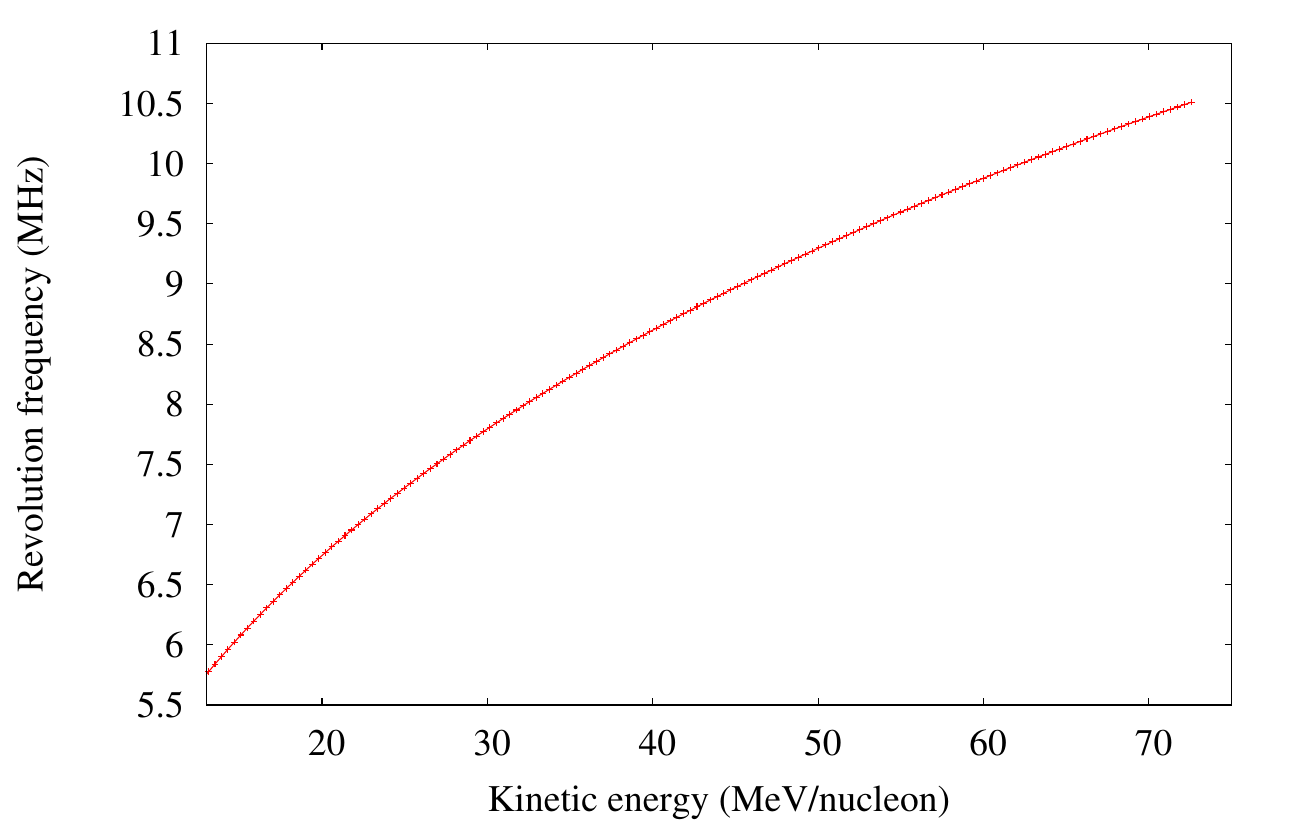}
\caption{Revolution frequency versus the kinetic energy per nucleon.}
\label{Revo_freq}
\end{figure}
The latter covers a range of 5.6-10.5 MHz as shown in fig \ref{Revo_freq}. Assuming a total energy gain of \mbox{2 keV/turn} results in an acceleration cycle that lasts 3.6 ms. If one assumes an equivalent period for the rise and fall of the RF, then the goal is to achieve a repetition rate of \mbox{150 Hz.}\\
The momentum phase slip factor at extraction is 0.64 corresponding to a transition (kinetic) energy of 1 GeV/nucleon. This ensures there is no crossing of such an energy for which the longitudinal phase stability vanishes \cite{kolomensky}. \\
The accelerated particle orbit at the location of the straights is shown in fig \ref{accel_or} where one can see that the \mbox{on-momentum} particle oscillates at $\pm 300 \text{ $\mu$m}$ around the ideal closed orbit. Given the slow acceleration rate (2keV/turn), this is in agreement with the dispersion function calculations predicting the orbital radius after i+1 turns: $$R(i+1) = R_0 + \sum_{k=0}^{i} D_x(p[k]) \dfrac{p[k+1]-p[k]}{p[k]}$$ The closed orbit distortion is substantially reduced thanks to the oscillatory behavior of the dispersion function with the energy.
 \\
Next, one evaluates the DA of the accelerator in presence of misalignment errors. The DA is defined here as the maximum initial horizontal normalized amplitude that the particle can have without any losses due to single particle dynamics effects over 30000 turns of the entire acceleration cycle. This is a fundamental criterion in order to determine the tolerance of the lattice to imperfections. Each of the 12 magnets is transversely (horizontally and vertically) offset randomly using a Gaussian distribution with a standard deviation $\sigma$ and a cutoff at $3\sigma$. For each error, 100 different patterns are tested and an initial normalized vertical displacement at 20 mm.mrad is chosen. The effect of the misalignment on the DA is shown in Fig. \ref{da} where one can see that with reasonable alignment errors ($\sigma \approx 100 \text{ $\mu$m}$) the DA is essentially unchanged. However, the DA of the lattice with trim quadrupoles is almost 3 times larger than the DA of the bare lattice. This is mainly due to the third integer resonance crossing problem with the bare lattice. As expected the lattice with the optimized working point has the largest DA. 

\begin{figure}[htb]
\centering 
\includegraphics*[width=8cm]{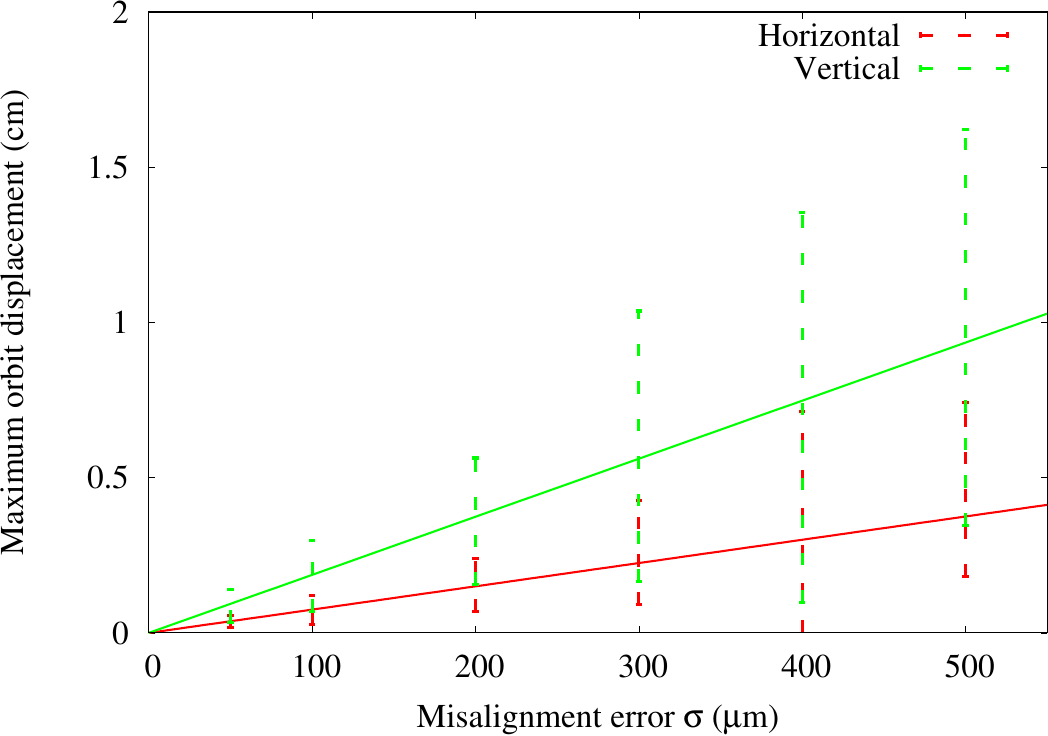}
\caption{The horizontal and vertical RMS accelerated orbit distortion with various misalignment errors: the solid lines show the minimum chi-squared fit to the data while the dashed lines show the RMS value of the maximum orbit distortion. The mean amplification factors for the orbit distortion (obtained from the fitted data) in the horizontal and vertical planes are 7.5 and 18.7 respectively.}
\label{cod}
\end{figure}

\begin{figure}%
\centering
\subfigure[][Acceleration without the time-varying field of the quadrupoles. 40 \% of the beam is lost at this stage.]{%
\label{fig}%
\includegraphics[height=2in]{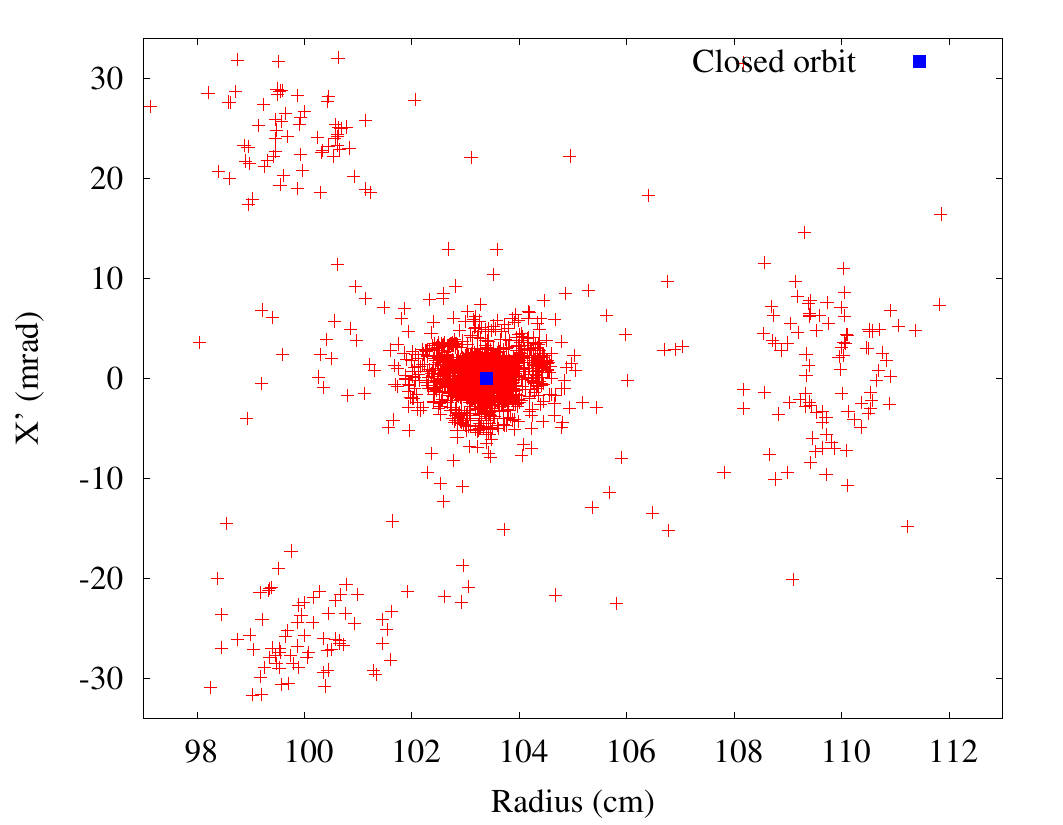}}%
\hspace{8pt}%
\subfigure[][Acceleration with the time-varying field of the quadrupoles. No occurring beam losses.]{%
\label{fig}%
\includegraphics[height=2in]{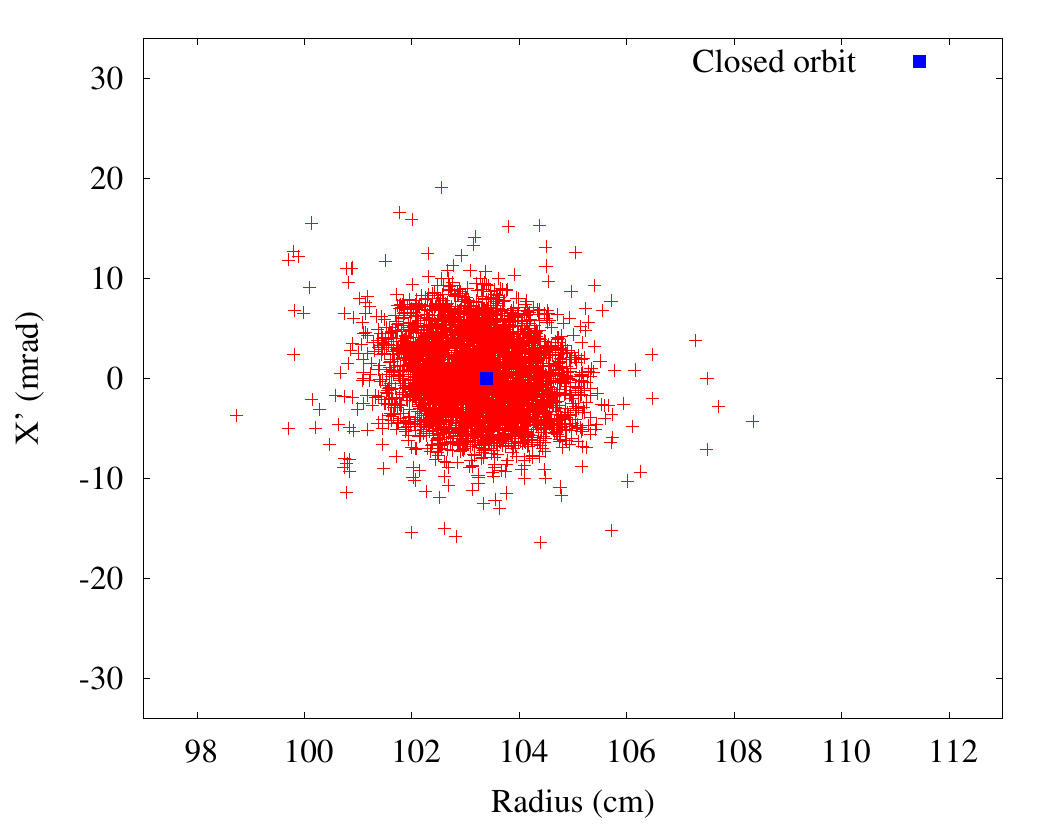}}%
\caption[A set of four subfigure.]
{Horizontal phase space distribution at 20 MeV/nucleon at the center of the long straight sections, 4000 turns tracked from injection.
}%
\label{fig:ex3}%
\end{figure}

Since one of the main objectives of this design is to achieve a variable energy extraction, one decided to calculate the maximum deviation of the accelerated orbit from the nominal one at the location of the extraction device during the entire acceleration cycle and in presence of misalignment errors as described above. In general, the maximum is reached near the injection cycle due to the adiabatic damping of the amplitude of particle oscillations with the momentum. The accelerated maximum rms orbit distortion for various misalignment errors is shown in fig \ref{cod}.
It can be seen that the vertical plane is the limiting problem for the variable energy extraction. However, this is still acceptable with a reasonable alignment error of the order of $\sigma \approx 100 \mu m$ corresponding to a maximum average orbit distortion of 1.8 mm. \\
Finally the impact of the trim quadrupoles on the beam transmission is illustrated in fig \ref{fig:ex3}. An injected matched distribution is tracked for 4000 turns from injection, and one can see that the beam losses are overcome due to the optimization of the phase space trajectories with the trim quadrupoles.

\section{Conclusion}
In this paper, a novel concept of compact racetrack FFA with dispersion free straight sections was proposed. It was shown that this is of particular interest for a heavy ion accelerator with variable energy extraction. In addition, one demonstrated for the first time that it is possible to overcome the problem of resonance crossings by time-varying the field of the trim quadrupoles during the acceleration cycle. This is of particular interest for high intensity applications and is a step further towards designing a tunable FFA with minimum beam losses. The study of the dynamics of this concept were investigated. 
In addition, the sensitivity to random misalignment errors of the magnets was explored and it was found that the mean DA is almost insensitive to the alignment error.

\end{document}